# Massive Antenna Arrays with Low Front-End Hardware Complexity: An Enabling Technology for the Emerging Small Cell and Distributed Network Architectures

Vlasis I. Barousis[1], Mohammad A. Sedaghat[2], Ralf R. Müller[3] and Constantinos B. Papadias[1]

*Abstract* - **This paper presents the current state-of-the-art of massive antenna array architectures with significant front-end hardware savings, as an enabler for future small and powerful cell nodes that will be able to carry massive MIMO technology. Radio frequency (RF) hardware architectures with a single power amplifier are reviewed, compared, and found superior to conventional MIMO implementations in terms of cost, dissipated heat, and physical size. This progress on the RF-side allows to merge the two competing cellular concepts of virtual and massive MIMO into a hybrid approach of remote radio heads with massive MIMO arrays.**

## I.    INTRODUCTION

To satisfy the continuously increasing demand for higher data rates and mobility and to meet the emerging users' expectations, industrial partners and operators should prepare for major changes ahead. The existing cellular network architectures seem to reach gradually their absolute limits, implying that the advent of new technologies and the advantageous exploitation of additional resources, e.g. more spectrum, is mandatory. Toward this goal, two are the main emerging visions of future cellular architectures.

Current technological advances propose the concept of *distributed base stations* that it is expected to replace the existing bulkier BS units and play a crucial role in the near future. A key enabler in this new approach is the so-called remote radio head (RRH) [1], which is a compact and lightweight radio module that implements advanced wireless standards such as Wimax and LTE. Besides the radio interface, RRHs also carry an appropriate optical interface to allow connections to the fixed backbone network via optical fibre backhauling. Under this new network perspective, intelligence is pushed away from the bulky BSs and aggregated toward the backhaul where central processing units handle the fast coordination/cooperation across a large number of cells. Moreover, the bulky BSs are replaced by smaller and lightweight radio nodes that are distributed in space and allow for widely geographically distributed access via radio-over-fibre connections to the central node or BS.

---

[1] Broadband & Wireless Sensor Networks Lab (B-WiSE), Athens Information Technology, Peania, Greece, email: {vbar, cpap}@ait.gr
[2] Norwegian University of Science and Technology, Trondheim, Norway, email: mohammad.sedaghat@iet.ntnu.no
[3] Friedrich-Alexander Universität Erlangen-Nürnberg, Erlangen, Germany, email: ralf@iet.ntnu.no

In contrast with this vision, a competitive approach called massive MIMO [2], [3] is based on a sparser but *centralized* network infrastructure where a large amount of radio elements and processing is conferred to fewer but larger in size BSs. The more powerful BSs are able to eliminate inter-cell interference by focusing the signal power closely around the intended receiver in both angular and radial domain, thus eliminating the need for additional cell-to-cell coordination and interference balancing/mitigation techniques that would increase significantly the network's burden. In view of the advantages of this alternative (more details can be found in [3]), certainly massive MIMO can increase substantially the capacity as a result of the very high multiplexing gain that potentially is available. Simultaneously, a significant increase of the radiation efficiency can be achieved that mainly results from the high beamforming gain and the ability to target the transmitted power to a small region in space, i.e. the intended receiver, only. This in turn implies that interference mitigation is inherent in those systems and no additional provision is necessary. Besides the apparent radiation benefits, massive MIMO can be built with *inexpensive* and *low-powered* radio-frequency (RF) circuit components. Indeed, the expensive and ultra-linear power amplifiers can be replaced by hundreds of low-cost amplifiers with output power in the milli-watt range. Although the requirements for highly linear behavior is relaxed in this case, all consequent side-effects are absorbed *since eventually what matters is their combined action*. Another benefit of this approach is the significant reduction of the latency on the air interface, etc. [3]. On the other hand, unfortunately the wide deployment of this approach is hampered by several limiting factors. Due to the large number of antenna elements, channel estimation becomes a non-trivial issue. It seems to be feasible only if the system is operated in time division-duplex mode (TDD) relying on the sensitive issue of channel reciprocity [3]. Pilot contamination, i.e. the lack of enough signal dimensions to fit orthogonal pilot sequences for channel estimation [3] still requires further research. Recent promising progress can be found e.g. in [4], [5].

It is understood that both aforementioned competing approaches aim to tide over the existing capacity limitations and propose future cellular architectures that will offer significant sum-capacity gains. In view of the emerging network demands and recognizing the pros and cons of the two aforementioned alternatives, in this paper we envision a balanced hybrid combination as it is illustrated in Fig. 1, i.e. a highly distributed BS architecture in which the intelligence and all necessary processing is centralized to a main node and multiple and more powerful RRHs, each carrying a massive array. Bearing in mind the strict requirements of RRHs, *the circuit front-end hardware complexity of the massive array structures emerges as the key enabler technology* that *will bring into reality this hybrid approach*. As shown in Fig. 1, each RRH could serve its dedicated area, e.g. a femtocell, and could be equipped with a massive array with low hardware front-end complexity, which could be driven by the central unit through optical links. The RRHs can be deployed for example on a rooftop, on a tower or they could be mounted on a wall.

Assuming a *unified* antenna array approach, in the following the main focus is on the review of novel front-end circuit architectures that allow for significant hardware savings and are appropriate to drive massive antenna arrays. In particular, Section II presents in a unified way the case of an arbitrary antenna architecture, as well as two special cases that lead to low hardware complexity front-ends. As a clear step forward, Section III in turn focuses on a novel front-end architecture that requires only a single carrier feeding and tunable loads and is best suited for massive array implementations.

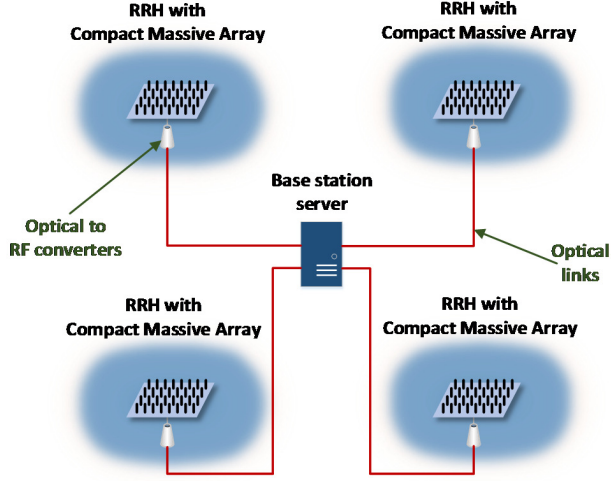

*Fig. 1: Proposed hybrid architecture. Each RRH can serve a femtocell and all RRHs are aggregated to a common central unit that holds the intelligence.*

## II. A Unified Treatment of Antenna Arrays and Front-End Hardware Complexity Architectures

### A. The arbitrary array architecture

This Section provides a unified and qualitative description of the front-end feeding architectures for antenna arrays, with a special focus on those that offer significant hardware savings. The main concept is illustrated in Fig. 2, where the arrays are considered in transmitting mode. In the arbitrary case shown in Fig. 2a , the $N$ antenna elements are deployed arbitrarily in space and therefore the array can be modeled by an arbitrary coupling matrix, whose diagonal entries $Z_{nn}, n = 1, 2, \ldots, N$ represent the self-impedances and all remaining $Z_{nm}, n, m = 1, 2, \ldots, N, n \neq m$ represent the mutual electromagnetic couplings between the $n$-th and $m$-th elements. Moreover, the $n$-th element is driven by its own RF chain through an arbitrary and fixed impedance of $Z_n$, which could model the output impedance of the front-end RF circuit block in the RF chain. Then, the currents at the port of the radiating elements are given by a generalized version of Ohm's law as:

$$\mathbf{i} = \left[ \mathbf{Z} + \underbrace{diag\left( Z_1, Z_2, \ldots, Z_N \right)}_{\text{diagonal loading matrix}} \right]^{-1} \underbrace{\left[ v_1, v_2, \ldots, v_N \right]}_{\text{feeding vector}}^{T} \tag{1}$$

### B. The Ideal Array Architecture

It should be worthy emphasized that the ideal assumption of uncoupled antenna elements that has been adopted widely in the literature of MIMO systems is another special case, as shown in Fig. 2b. As observed, the effective coupling $\mathbf{Z} + diag\left( Z_1, Z_2, \ldots, Z_N \right)$ in (1) becomes diagonal and hence the currents at the ports of the elements and the triggering voltages are a scaled version of each other. This justifies the common assumption in the relevant MIMO literature that the voltages are assumed as the input to the MIMO system. However, as shown in (1) and explained very recently in [6], in order to capture the effects of any analog front-end feeding circuitry, the currents are the correct input quantities.

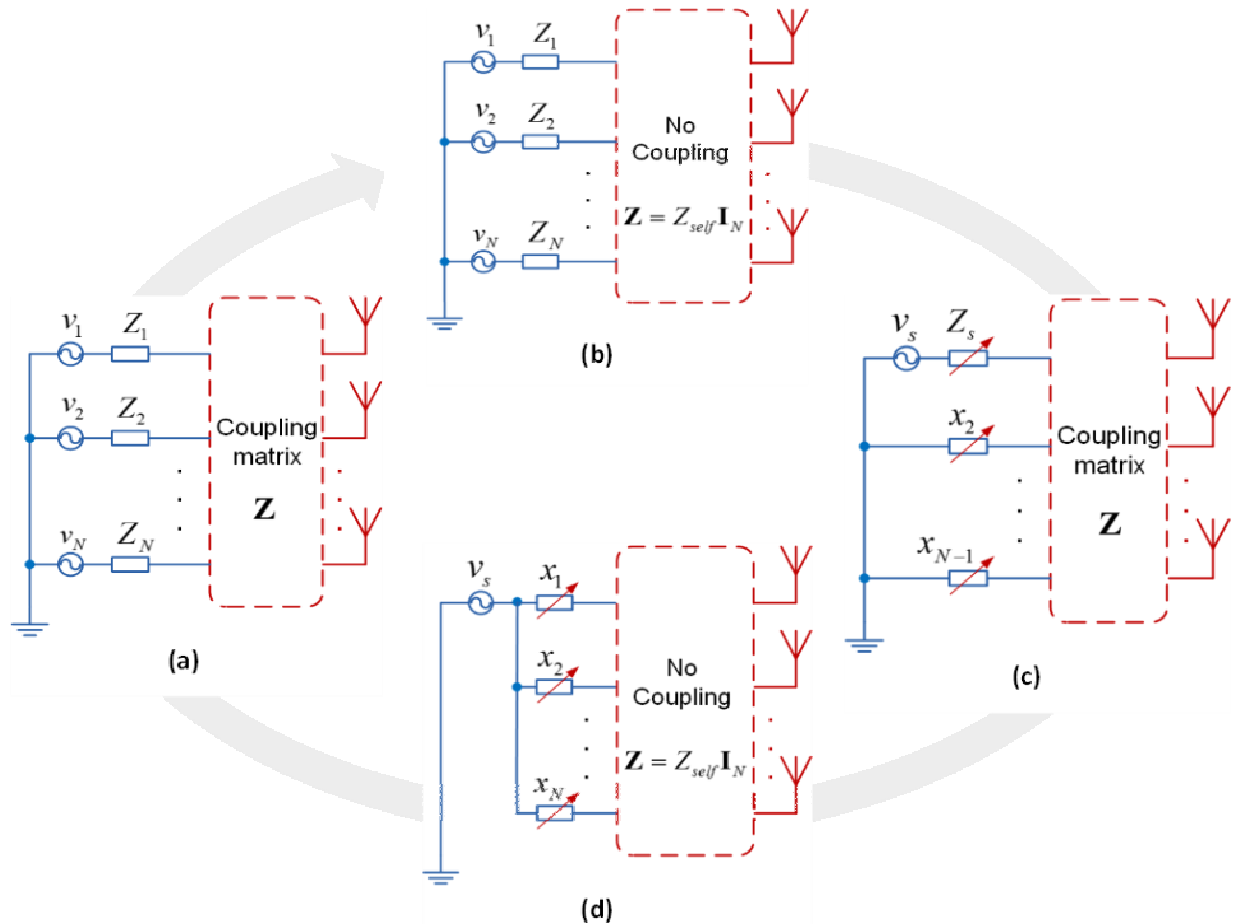

*Fig. 2: Unified illustration of different front-end circuit feeding architectures for antenna arrays.*

### C. Parasitic Array Architectures

An attempt to achieve hardware savings is drawn in Fig. 2c that describes the principle of operation of a special antenna structure called *parasitic array*, in which only a single *active element* is fed by an external RF chain while the remaining ones are terminated through tunable analog loads and therefore are known as *parasitics*. Those arrays were first introduced by Harrington [7] in 1978, but they gained attention more recently by Gyoda and Ohira who proposed their use for low-cost analog beamforming applications [8]. In contrast to the conventional case, their functionality is essentially based on the strict requirement for strong mutual coupling among all elements. This guarantees that the sole feeding at the active element can induce adequate currents to the ports of all parasitics and hence all elements participate to the shaping of the total radiation pattern. The input currents can be further controlled by changing the effective couplings, which in turn is attained by tuning the analog loads. In this way, beamforming can be achieved. The input currents to all elements are still given by (1), substituting $Z_n = x_n$, $n = 2, 3, \ldots, N$ and $v_n = 0$, $n = 2, 3, \ldots, N$. As observed, this approach offers significant hardware savings, as the number of RF chains is reduced to a single one and the remaining RF chains are replaced by easy-to-implement tunable analog loads. Moreover, it is noted that the variable $Z_s$ models the combined effect of the output impedance of the RF chain and an arbitrary tunable matching circuit that follows the arrays' input impedance, which in turn is a function of the variable loads and the couplings.

Thanks to the tunable analog loads and the consequent beam-shaping abilities, parasitic arrays have been recently proposed for *single RF MIMO systems*. The story begins in the pioneering work in [9] where the authors propose a smart analog switching technique in order to multiplex two symbols over the air starting from the simple case of on-off keying (OOK) and continuing to low order PSK modulation formats. This introductory work laid the foundation for further research on single RF MIMO transmission with parasitic arrays and a detailed overview can be found in [10], where among others a proof-of-concept experiment is described that validates their functionality. These works assume purely imaginary loads and succeed to multiplex any PSK signals over the air. To support constellations beyond PSK the authors in [11] propose an *active circuit design* with appropriately biased field effect transistors (FETs), which implements a tunable complex loading for parasitic elements with the real, i.e. resistive, part ranging from negative to positive values. As a clear step forward to design compact and lightweight parasitic arrays, [12] provides the design guidelines and specifications that should be met in order to achieve an arbitrary space-time precoding scheme. *The common baseline in all these works is to design appropriate antenna arrays and loading circuits and emulate the MIMO effect in the air by switching to different transmit radiation patterns in consecutive signaling periods*.

### D. The Load-Modulated Array Architecture

Inspired by the load tuning as a means to control the array's response, Fig. 2d illustrates a newly introduced approach, whose main competitive advantage is *that it completely eliminates the need for any RF chain while a single, unmodulated carrier signal suffices for a proper functionality*. As observed, all elements are connected to a common carrier signal source through passive and lossless two-port loading networks, each implemented with tunable reactance elements. The purpose of those tunable loads is to adjust the input currents to all radiating elements, *thereby implementing a desired signal constellation in the analog domain*. This alternative solution is known as *load-modulated arrays* [13], [14] in order to reflect their principle of operation. The input currents to the antenna elements are given again by (1), but by replacing the entries of the diagonal impedance matrix by the output impedance of each loading 2-port network and the entries of the voltage vector by the Thenevin equivalent voltages, as seen by the output of the 2-port networks. It is noted that since all elements are fed by a common (but at the end independently modified) external signal, couplings among the elements are not necessary and therefore an uncoupled array is shown as it has been originally reported in the literature. However, the case of a compact and at some extend coupled arrays should not be excluded.

### III.  MASSIVE ARRAY ARCHITECTURES WITH LOW FRONT-END HARDWARE COMPLEXITY

### A. Front-End Architecture and Challenges

Although the availability of digital circuitry has been following Moore's law for long, the cost of analog hardware has remained stable. Although (as briefly explained in Section I) massive MIMO is a promising technology with proven and remarkable benefits, their wide deployment is hampered mainly by the highly increased hardware complexity of the front-end RF circuits. Considering a conventional large array with independently fed antenna elements, the total cost and the hardware burden grows linearly with the number of RF chains and with the number of RF power amplifiers in particular. The requirement for highly linear characteristics and for reducing the power dissipation of power amplifiers is another crucial problem. In particular, conventional multi-fed MIMO systems often require a high peak-to-average power ratio (PAPR) that might reach the level of 8 dB to 12 dB [14] e.g. in orthogonal frequency-division multiplexing (OFDM) modulation and/or linear precoding. Therefore, it turns out that in the conventional case the use of highly linear RF power amplifiers is mandatory in order to avoid possible distortion of the output signal and generation of harmonic frequencies, which inevitably will reflect to system's performance degradation. Unfortunately, the high linearity of an amplifier is always tied to low efficiency. Indeed, in this case the active device, e.g. a bipolar or FET transistor, should be biased so that to conduct during the entire cycle of the input signal. This in turn will reduce largely the efficiency of the entire RF block due to the high quiescent current (for example the efficiency of a class A power amplifier often is upper bounded to 25 %). From circuit design point of view, switching power amplifiers comprised by a symmetric topology of complementary blocks (e.g. push-pull) and multiple active elements can be used to increase the RF efficiency. Hence, each active element and the accompanied circuit block will conduct only during small portions of the input signal's cycle. Although the switching mode improves the total efficiency, this comes at the cost of linearity, caused e.g. due to crossover distortion effects. Based on the above it is easily understood that *in conventional MIMO systems operators pay for high peak power and not for the average power*. Definitely, the RF front-end complexity and cost is the main limiting factor that restricts the number of elements in massive MIMO systems and thus the overall potential performance of cellular networks.

*B.  Proposed Massive Array Architecture and Indicative Performance Results*

The novel architecture in Fig. 2c constitutes the enabling technology that will bridge that gap and will provide low cost massive array implementations with significant hardware savings, which will bring the envisioned hybrid network topology in Fig. 1 closer to reality. As shown in more detail in Fig. 3 and explained in Section I, all elements are connected to the sole power amplifier through passive, two-port lossless loading networks of "T" or "Π" topology [13], whose purpose is to adjust the input currents to the radiating elements based on the desired signaling format. Unlike the parasitic arrays, the advantageous point is that all elements are connected to a common source that feeds a fixed sinusoid carrier signal. A circulator is also assumed that protects the power amplifier against the reflected power, which (if any) will be consumed onto the resistance $R$ [13], [14]. *Packing a large number of antenna elements and assuming a massive array implementation the proposed front-end architecture becomes more favorable and the hardware drawbacks tend to eliminate, as explained in the following*.

*The PAPR tends asymptotically to one*

For load-modulated arrays with a common power amplifier as shown in Fig. 3, the PAPR is greatly reduced to one as the number of antenna elements becomes larger. This is explained by the law of large numbers. In fact, it results from the averaging effect which dominates when many antenna elements co-exist. Assuming complex Gaussian input data signals and $N$ antenna elements, the authors in [13] explain that the output power is $\chi^2$- distributed with $2N$ degrees of freedom. The reduction of the crest factor, i.e. the square root of the PAPR value, due to the averaging is drawn in Fig. 4a for different values of clipping probability. As observed, for $N = 100$ elements and 0.1 % clipping probability the crest factor is found around 1.2 dB. Referring to the discussion in Section III.A, *this definitely indicates that there is no need for high linearity and the use of low-efficient class A/B power amplifiers. Instead, a single and highly-efficient class F power amplifier can be used that allows for smaller linear range but offers significantly higher efficiency of around 80% [15], which is attained due to its switching mode of operation*.

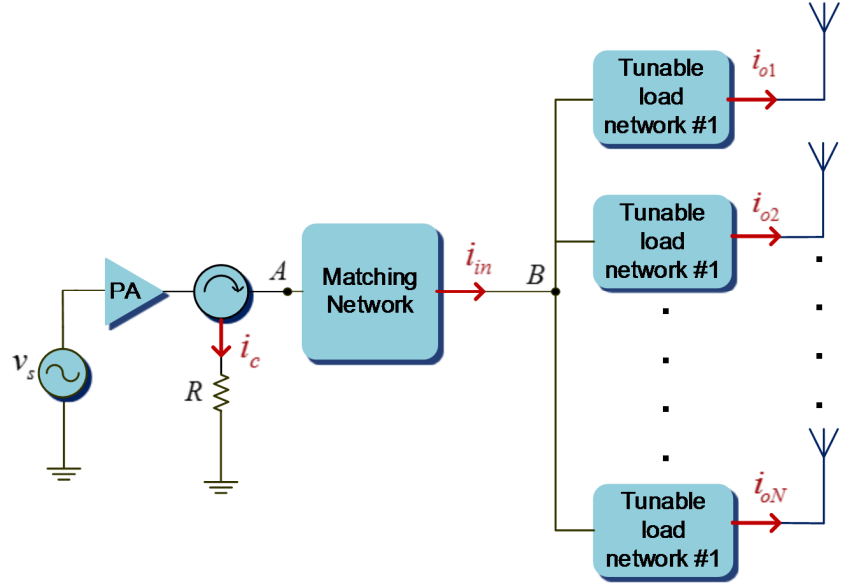

*Fig. 3: Load-modulated arrays for low cost massive MIMO design with low hardware complexity and high efficiency.*

### *Good matching conditions are achieved*

As opposed to conventional MIMO transmitters that are equipped with *fixed* RF front-end modules and externally varying voltages, Fig. 3 shows that load modulated arrays are driven by a fixed carrier signal source and the currents $i_{o1} \ldots i_{0N}$ that implement a desired signaling format are controlled indirectly by tuning the front-end circuits. Therefore, it is evident that *the input impedance seen from the analog source's side varies with time as a function of the desired currents $i_{o1} \ldots i_{0N}$ in consecutive signaling periods.*

Thankfully, as the number of elements grows and the massive regime is reached, the load modulation tends to be insensitive to mismatch effects. This is illustrated in Fig. 4b that draws the probability density function (pdf) of the voltage standing wave ration (VSWR) for a growing number of antenna elements. The trend that is clearly observed is that $\lim_{N \to \infty} VSWR = 1$, which implies that as the number of elements grows the VSWR converges to 1 and therefore the reflection coefficient converges to zero. In turn, the input impedance of the load-modulated arrays becomes purely Ohmic, i.e. real, and nearly equal to the output resistance of the source.

### *Low signal distortion*

In conventional systems, the diverse input signals are clipped independently, as a result of the existence of individual RF chains to the antenna elements. On the contrary, in load-modulated arrays the input signals are clipped only if the desired sum power becomes higher than the one that can be provided externally by the sole power amplifier. The authors in [14] analyze two strategies, i.e. *clipping with respect to the minimum mean squared error (MMSE)* and *equal clipping of all signals*. The first approach minimizes the expression of $\min_{\hat{i}_{on}} \sum_{n=1}^{N} \left( i_{on} - \hat{i}_{on} \right)^2$, where $\hat{i}_{on}$ stands for the clipped version of the desired current $i_{on}$ at the $n$-th antenna element. The second approach clips all signals equally in order to satisfy the source's power constraints.

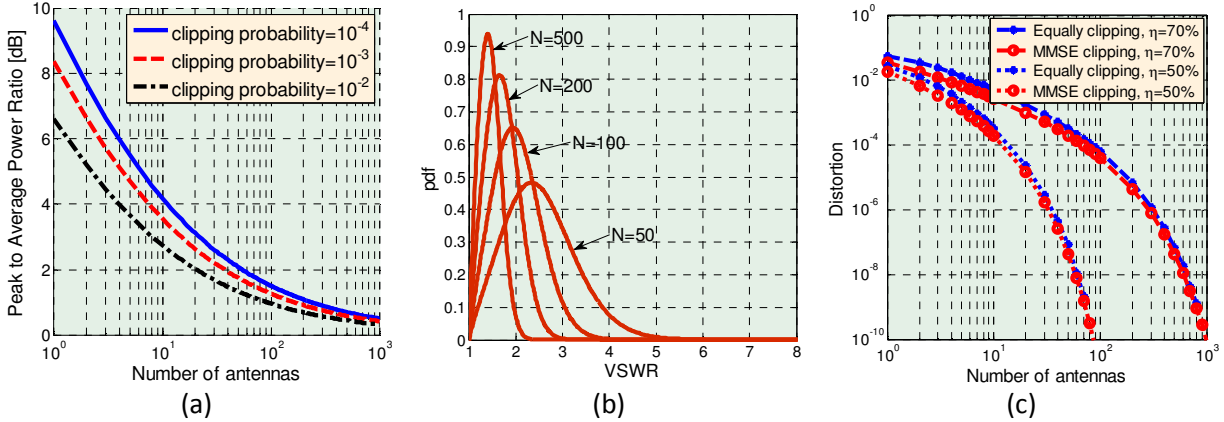

*Fig. 4: a) PAPR vs. number of antennas for different clipping probabilities, b) VSWR distribution for different numbe of antenna elements and c) signal distortion vs. number of antennas for different clipping methods and total efficiencies.*

Obviously, the clipping effect introduces a signal distortion that thankfully weakens as the number of antennas grows. In fact, clipping in single-RF MIMO transmitters is less harmful than in conventional MIMO transmitters. Indicatively, Fig. 4c draws the distortion versus the number of antennas for fixed values of total power efficiency. It is observed that the two clipping exhibit almost the same behavior and the distortion becomes negligible for number of elements beyond 100.

## C. Implementation of the Analog Loads

As it has been shown already, the hardware complexity scales mainly with the number of RF chains and especially with the total RF peak power. Therefore, savings have been attained by replacing the bulkier RF chains by simple-to-implement analog circuits with tuning capabilities. Tunable loads were first considered in parasitic arrays, e.g. see [10], where a tunable capacitor was realized by tuning the reverse bias voltage of varactor diodes. As a straightforward extension, "T" or "Π" two-port network topologies can be assumed in both parasitic arrays and load-modulated arrays in which a combination of inductors and tunable capacitances could offer advanced tuning capabilities. As an extension to this approach, a six-port analog modulator has been demonstrated in [16] that uses Schottky diodes and achieves high switching speed, enhancing the achievable data rate up to 1.2 Gbit/second.

An alternative implementation is by networks of pin-diodes. With 64k integrated pin-diode switches, one out of $2^{16}$ predefined loads can be chosen. Given the fast decay of digital hardware costs, even higher resolution that 16 bit accuracy seems possible. Furthermore, accuracy can be improved by shaping of the quantization noise and passive post-processing by means of surface acoustic wave (SAW) filters. This fully digital implementation has the advantage that is does not require any digital-to-analog (D/A) converters. However, a power loss around 1 or 2 dB caused by the SAW filters is to be tolerated.

## IV.   CONCLUSIONS & OUTLOOK

Hardware savings in MIMO transceivers has been a crucial and challenging problem. Massive arrays gain an increasing attention and tend to be one of the leading technologies that will be involved in the emerging 5G HetNets / small cells network architectures. The perceived interest to unlicensed bands in higher microwave and even milli-meter wave regimes will make massive arrays inevitable to overcome the path loss. Furthermore, the smaller wavelengths will aid their inclusion into size-restricted access points. This paper promotes a newly appeared front-end RF architecture that offers significant hardware savings and is best suited for massive array deployments. This low-complexity RF-architecture will be the key enabling technology to pave the way for novel and emerging cellular architectures, as shown in the hybrid approach in Fig. 1. The proposed massive array architecture needs only a single carrier feeding signal, thus eliminating completely the need for a complete RF chain. With digital load modulators, it even eliminates the need for D/A converters. Under this perspective, the input currents to the diverse antenna elements are adjusted by tuning the front-end feeding circuit itself, which in turn is attained by varying analog loads that are attached to all antenna branches. Letting the number of elements increase, it has been shown that a switching power amplifier suffices and allows for significant power savings. Moreover, the mismatch effects as well as the signal distortion tend to attenuate and become negligible. Overall, this paper provides a roadmap for further exploration towards this direction and aims to trigger for more intensive research for future lightweight, powerful massive MIMO transceivers that will bring flexible cellular networks with powerful nodes closer to reality.

REFERENCES

[1] G. Kardaras and C. F. Lanzani, "Advanced multimode radio for wireless and mobile broadband communication," *In Proc. IEEE European Wireless Technology Conference (EuWiT)*, 2009.

[2] F. Rusek, D. Persson, B. K. Lau, E. G. Larsson, T. L. Marzetta, O. Edfors and F. Tufvesson, "Scaling up MIMO: Opportunities and challenges with very large arrays*," IEEE Signal Processing Magazine*, vol. 30, no. 1, pp. 40-60, Jan. 3013.

[3] E. Larsson, O. Edfors, F. Tufvesson and T. Marzetta, "Massive MIMO for next generation wireless systems," *IEEE Communications Magazine*, vol.52, no.2, pp.186,195, February 2014.

[4] Y. Haifan, D. Gesbert, M. C. Filippou and L. Yingzhuang, "Decontaminating pilots in massive MIMO systems," In Proc. IEEE International Conference on Communications (ICC), pp. 3170-3175, 9-13 Jun. 2013.

[5] R.R. Müller, L. Cottatellucci, M. Vehkaperä, "Blind Pilot Decontamination," *IEEE Journal of Selected Topics in Signal Processing, vol. 8, no. 10, Oct. 2014.*

[6] V. I. Barousis, C. B. Papadias and R. R. Müller, "A new signal model for MIMO communication with compact parasitic arrays," In Proc. IEEE Communications, Control and Signal Processing, Athens, Greece May 21th-23th, 2014.

[7] R. Harrington, "Reactively controlled directive arrays," *IEEE Transactions on Antennas and Propagation*, vol. 26, no. 3, pp. 390–395, 1978.

[8] T. Ohira and K. Gyoda, "Electronically steerable passive array radiator antennas for low-cost analog adaptive beamforming," in In Proc. IEEE Phased Array Systems and Technology, pp. 101–104, 2000.

[9] A. Kalis, A. Kanatas, and C. Papadias, "A novel approach to MIMO transmission using a single RF front end," IEEE Journal on Selected Areas in Communications, vol. 26, no. 6, pp. 972–980, 2008.

[10] A. Kalis, A. Kanatas, and C. Papadias, Parasitic Antenna Arrays for Wireless MIMO Systems. Springer, New York, USA, 2013.

[11] B. Han, V. Barousis, A. Kalis, C. Papadias, A. G. Kanatas and R. Prasad, "A Single RF MIMO Loading Network for High Order Modulation Schemes," International Journal on Antennas and Propagation, Vol. 2014.

[12] V. I. Barousis and C. B. Papadias, "Arbitrary Precoding with Single-Fed Parasitic Arrays: Closed-Form Expressions and Design Guidelines," *IEEE Wireless Communications Letters,* vol.3, no.2, pp.229-232, April 2014.

[13] R. R. Müller, M. A. Sedaghat and G. Fischer, "Load modulated MIMO," Communication Theory Workshop, May 25-28, 2014, Curaçao,

[14] M. A. Sedaghat, R. R. Müller and G. Fischer, "A novel single-RF transmitter for massive MIMO," In Proc. ITG Workshop on Smart Antennas, Mar. 2014, Erlangen, Germany.

[15] S.C. Cripps, RF Power Amplifiers for Wireless Communications, Norwood, MA: Artech House, 2006.

[16] J. Östh, Owais, M. Karlson, A. Serban and S. Gong, "Schottky diode as high-speed variable impedance load in six-port modulators," In Proc. IEEE International Conference on Ultrawideband, Sep. 2011, Bologna, Italy.